\pdfoutput=1

\documentclass[11pt]{article}

\usepackage[]{EMNLP2022}

\usepackage{times}
\usepackage{latexsym}
\usepackage[T1]{fontenc}
\usepackage[utf8]{inputenc}
\usepackage{microtype}
\usepackage{inconsolata}

\usepackage{bm}
\usepackage{subfigure}
\usepackage{multirow}
\usepackage{arydshln} 
\usepackage{enumitem}

\usepackage{amsmath}
\usepackage{amsfonts}
\usepackage{algorithmic}
\usepackage{graphicx}
\usepackage{textcomp}
\usepackage{xcolor}
\usepackage[utf8]{inputenc}
\usepackage{CJK}
\usepackage{comment}
\usepackage{float}
\usepackage{algorithm}  
\usepackage{mathrsfs}
\usepackage{makecell}
\usepackage{multirow}

\usepackage{soul}
\usepackage{textcomp} 
\usepackage{booktabs}
\usepackage{mathrsfs}
\usepackage{etoolbox}
\usepackage{marvosym}
\usepackage{pifont}
\usepackage{hyperref}
\usepackage{amsbsy}
\usepackage{setspace}
\usepackage{amssymb}

\title{Knowledge Distillation based Contextual Relevance Matching \\
for E-commerce Product Search}

\author{Ziyang Liu\textsuperscript{\S},Chaokun Wang\textsuperscript{\S}\textsuperscript{*}, Hao Feng\textsuperscript{\S}, Lingfei Wu\textsuperscript{\dag}, Liqun Yang\textsuperscript{\ddag}\\
  \textsuperscript{\S}Tsinghua University,\textsuperscript{\dag}JD.com,\textsuperscript{\ddag}CNAEIT\\
  \textsuperscript{\S}\texttt{liu-zy21@mails.tsinghua.edu.cn,chaokun@tsinghua.edu.cn} \\
  \texttt{\textsuperscript{\dag}lwu@email.wm.edu, \textsuperscript{\ddag}yanglq@cnaeit.com} \\}

\begin{document}
\maketitle

\begingroup\renewcommand\thefootnote{*}
\footnotetext{Chaokun Wang is the corresponding author.}

\begin{abstract}
Online relevance matching is an essential task of e-commerce product search to boost the utility of search engines and ensure a smooth user experience.
Previous work adopts either classical relevance matching models or Transformer-style models to address it. However, they ignore the inherent bipartite graph structures that are ubiquitous in e-commerce product search logs and are too inefficient to deploy online.
In this paper, we design an efficient knowledge distillation framework for e-commerce relevance matching to integrate the respective advantages of Transformer-style models and classical relevance matching models. Especially for the core student model of the framework, we propose a novel method using $k$-order relevance modeling.
The experimental results on large-scale real-world data (the size is 6$\sim$174 million) show that the proposed method significantly improves the prediction accuracy in terms of human relevance judgment.
We deploy our method to the anonymous online search platform. The A/B testing results show that our method significantly improves 5.7\% of UV-value under price sort mode.
\end{abstract}

\section{Introduction}

\textbf{Relevance matching}~\cite{drmm, relevance_matching_semantic_matching, relevance_matching} is an important task in the field of ad-hoc information retrieval~\cite{ad_hoc_IR}, which aims to return a sequence of information resources related to a user query~\cite{IR1,IR2,emnlp-IR-1}. Generally, texts are the dominant form of the user query and returned information resources.
Given two sentences, the target of relevance matching is to estimate their relevance score and then judge whether they are relevant or not. However, text similarity does not mean semantic similarity. For example, while ``mac pro 1.7GHz'' and ``mac lipstick 1.7ml'' look alike, they describe two different and irrelevant products.
Therefore, relevance matching is important, especially for long-term user satisfaction of e-commerce search~\cite{ecommerce-6, ecommerce-5, emnlp-ecommerce-1, emnlp-ecommerce-2}.

\noindent \textbf{Related work.} With the rapid development of deep learning, the current research on relevance matching can be grouped into two camps (see Appendix~\ref{Related Work} for further details): 1.\underline{\textit{Classical Relevance Matching Models.}} For the given query and item, the classical relevance matching models either learn their individual embeddings or learn an overall embedding based on the calculation from word-level interaction to sentence-level interaction. The representative methods include ESIM~\cite{esim} and BERT2DNN~\cite{bert2dnn}.
2.\underline{\textit{Transformer-style Models.}} These models adopt the multi-layer Transformer network structure~\cite{transformer}. They have achieved breakthroughs on many NLP tasks and even reached human-level accuracy. The representative methods include BERT~\cite{bert} and ERNIE~\cite{ernie}. 

Although Transformer-style models show satisfactory performance on relevance matching, they are hard to deploy to the online environment due to their high time complexity.
Moreover, the above methods cannot deal with the abundant context information (i.e., the neighbor features in a query-item bipartite graph) in e-commerce product search.
Last but not least, when applied to real-world scenarios, existing classical relevance matching models directly use user behaviors as labeling information (Fig.~\ref{motivation}). However, this solution is not directly suitable for relevance matching because user behaviors are often noisy and deviate from relevance signals~\cite{clickmodel,clickmodel-2}.

\begin{figure}[t]
\centering
\setlength{\abovecaptionskip}{0.0cm}
\setlength{\belowcaptionskip}{-0.3cm}
\includegraphics[width=0.45\textwidth]{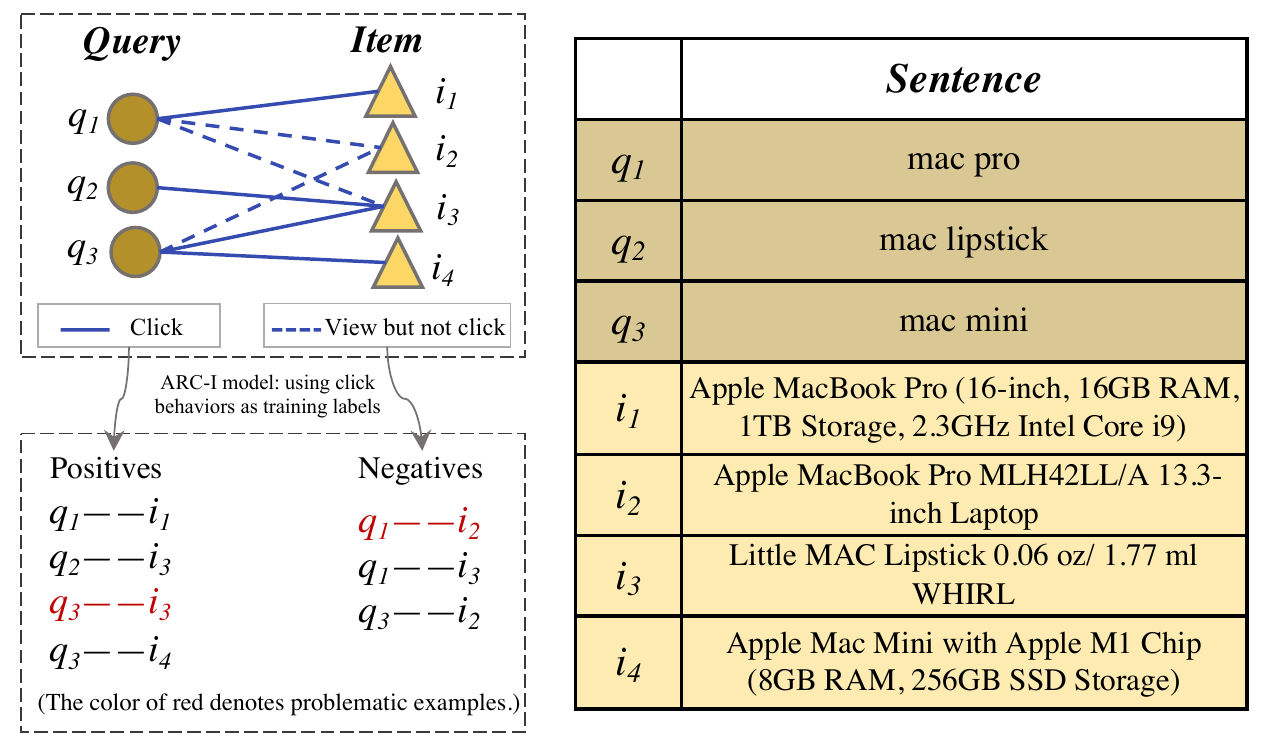}
\hspace{0in}
\caption{Shortcoming of the existing relevance matching model. Here we take the ARC-I model as an example. 
The right part shows the ground truth of queries and item titles.
The left part shows two problematic examples in ARC-I, which deviate from the ground truth.}
\vspace*{-2mm}
\label{motivation}
\end{figure}


In this paper, we propose to incorporate bipartite graph embedding into the knowledge distillation framework~\cite{emnlp-kg-1,emnlp-kg-2,emnlp-kg-3,emnlp-kg-4,kg-1} to solve the relevance matching problem in the scene of e-commerce product search. We adopt BERT~\cite{bert} as the teacher model in this framework.
We design a novel model called BERM, \textbf{B}ipartite graph \textbf{E}mbedding for \textbf{R}elevance \textbf{M}atching (BERM), which acts as the student model in our knowledge distillation framework. This model captures the 0-order relevance using a word interaction matrix attached with positional encoding and captures the higher-order relevance using the metapath embedding with graph attention scores.
For online deployment, it is further distilled into a tiny model BERM-O.

Our main contributions are as follows:
\begin{itemize}[leftmargin=*]
\vspace*{-3mm}
    \item We formalize the $k$-order relevance problem in a bipartite graph (Section~\ref{Problem Definition}) and address it by a knowledge distillation framework with a novel student model called BERM.
\vspace*{-3mm}
    \item We apply BERM to the e-commerce product search scene with abundant context information (Section~\ref{BERM Model}) and evaluate its performance (Section~\ref{experiments}). The results indicate that BERM outperforms the state-of-the-art methods.
\vspace*{-3mm}
    \item To facilitate online applications, we further distill BERM into a faster model, i.e., BERM-O. The results of online A/B testing indicate that BERM-O significantly improves 5.7\% (relative value) of UV-value under price sort mode. 
\end{itemize}

\section{Method}
\label{Method}

\subsection{Problem Definition}
\label{Problem Definition}

\begin{figure*}[t]
\centering
\includegraphics[width=0.7\textwidth]{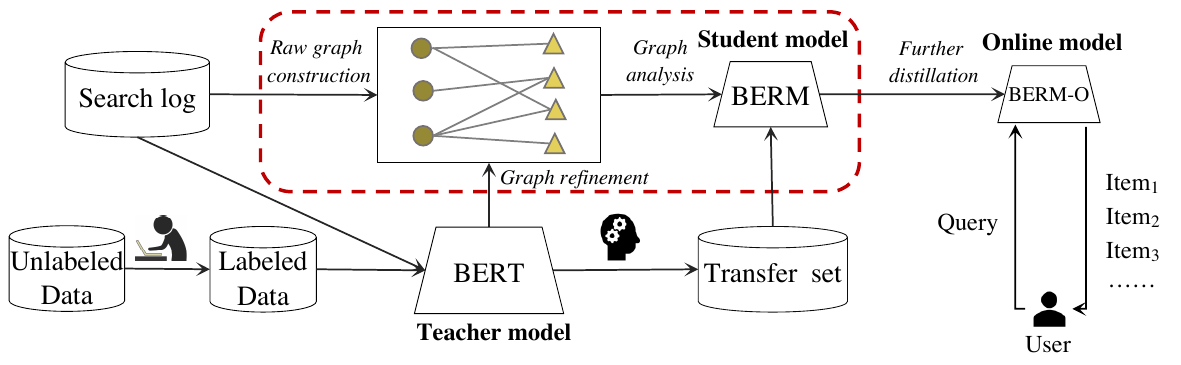}
\hspace{0in}
\setlength{\abovecaptionskip}{-0cm}
\setlength{\belowcaptionskip}{-0.4cm}
\caption{The e-commerce knowledge distillation framework proposed in our work. Three models are used in this framework: teacher model BERT, student model BERM, and online model BERM-O.}
\label{model1}
\end{figure*}

\newtheorem{myDef}{Definition}
\newtheorem{myTheo}{Theorem}
\newtheorem{myExm}{Example}
We first give the definition of the bipartite graph:
\begin{myDef}
    \textbf{Bipartite Graph}. Given a graph $\mathcal{G} = (\mathcal{U}, \mathcal{V}, \mathcal{E},$ $ \mathcal{A}, \mathcal{R})$, it contains two disjoint node sets $\mathcal{U}: \left \{u_{1}, u_{2}, \cdots , u_{n}\right \}$ and $\mathcal{V}: \left \{v_{1}, v_{2}, \cdots , v_{n'}\right \}$. For edge set $\mathcal{E}: \left \{e_{1}, \cdots , e_{m}\right \}$, each edge $e_i$ connects $u_j$ in $\mathcal{U}$ and $v_k$ in $\mathcal{V}$. In addition, there is a node type mapping function $f_{1}: \mathcal{U}\cup\mathcal{V} \rightarrow \mathcal{A}$ and an edge type mapping function $f_{2}: \mathcal{E}\rightarrow \mathcal{R}$. Such a graph $\mathcal{G}$ is called a bipartite graph.
\label{bipartite-graph}
\end{myDef}

\begin{myExm}
Given an e-commerce search log, we can build a query-item bipartite graph as shown in the left part of Fig.~\ref{motivation}, where $\mathcal{A} = \left \{Query, Item\right \}$ and $\mathcal{R}=\left \{Click \right \}$.
\end{myExm}

In a bipartite graph, we use the metapath and metapath instance to incorporate the neighboring node information into relevance matching. They are defined as follows:

\begin{myDef}
    \textbf{Metapath and Metapath Instance in Bipartite Graph}. Given a bipartite graph $\mathcal{G}=(\mathcal{U}, \mathcal{V}, \mathcal{E}, \mathcal{A}, \mathcal{R})$, a metapath $P_{i}=a_{1}\overset{r_{1}}{\rightarrow}a_{2}\overset{r_{2}}{\rightarrow} \cdots \overset{r_{l}}{\rightarrow}a_{l+1}$ ($a_{j} \neq a_{j+1}, 1 \leqslant  j \leqslant  l$) represents the path from $a_{1}$ to $a_{l+1}$ successively through $r_{1}, r_{2}, \cdots,r_{l}$ ($a_j\in\mathcal{A}$, $r_j\in\mathcal{R}$). 
    The length of $P_{i}$ is denoted as $\left | P_{i}\right |$ and $\left | P_{i}\right |=l$.
    For brevity, the set of all metapaths on $\mathcal{G}$ can be represented in regular expression as $P^{\mathcal{G}}=\left ( a|\varepsilon\right )\left ( a'a\right )^{+}\left ( a'|\varepsilon\right )$ where $a, a' \in \mathcal{A}$, $a\neq a'$. 
     The metapath instance $p$ is a  definite node sequence instantiated from metapath $P_{i}$. All instances of $P_{i}$ is denoted as $I(P_{i})$, then $p \in I(P_{i})$.
\end{myDef}

\begin{myExm}
    As shown in Fig.~\ref{motivation}, an instance of metapath ``$Query$-$Item$-$Query$'' is ``$q_{2}$-$i_{3}$-$q_{3}$''.
\end{myExm}

\begin{myDef}
\label{definition2}
    \textbf{$k$-order Relevance}. Given a bipartite graph $\mathcal{G} = (\mathcal{U}, \mathcal{V}, \mathcal{E}, \mathcal{A}, \mathcal{R})$, a function $F^{k}_{rel}: \mathcal{U}\times \mathcal{V} \rightarrow \left [ 0, 1\right ]$ is called a $k$-order relevance function on $\mathcal{G}$ if $F^{k}_{rel}(u_i,v_j)=G(\Phi (u_{i}), \Phi (v_{j}) | C_{k})$, where $\Phi(\cdot)$ is a function to map each node to a representation vector, $G(.)$ is the score function, $u_{i} \in \mathcal{U}$, $v_{j} \in \mathcal{V}$, and context information $C_{k} = \bigcup_{I_{P_{i}}\subseteq I(P_{i}), P_{i}\in P^{\mathcal{G}},\left | P_{i}\right |= k}I_{P_{i}}$.
\end{myDef}

Many existing relevance matching models~\cite{dssm, cdssm, ARC-I-ARC-II} ignore context information $C_{k}$ and only consider the sentences w.r.t.~the query and item to be matched, which corresponds to 0-order relevance (for more details, please see the ``Related Work'' part in Appendix~\ref{Related Work}).
We call it \emph{context-free relevance matching} in this paper. Considering that both the 0-order neighbor (i.e., the node itself) and $k$-order neighbor ($k \geqslant 1$) are necessary for relevance matching, we argue that a reasonable mechanism should ensure that they can cooperate with each other. Then the research objective of our work is defined as follows:

\begin{myDef}
\label{definition4}
    \textbf{Contextual Relevance Matching}. Given a bipartite graph $\mathcal{G} = (\mathcal{U}, \mathcal{V}, \mathcal{E}, \mathcal{A}, \mathcal{R})$, the task of contextual relevance matching is to determine the context information $C_{k}$ on $\mathcal{G}$ and learn the score function $G(\cdot)\;$.
\end{myDef}

\subsection{Overview}
\label{overview}
We propose a complete knowledge distillation framework (Fig.~\ref{model1}), whose student model incorporates the context information, for contextual relevance matching in e-commerce product search. The main components of this framework are described as follows (see Appendix~\ref{Details of the framework} for further details): 
\begin{itemize}[leftmargin=*]
    \vspace*{-3mm}
    \item \textbf{Graph construction.} We firstly construct a raw bipartite graph $\mathcal{G}$ based on the search data collected from the anonymous platform. Then we construct a knowledge-enhanced bipartite graph $\mathcal{G}'$ with the help of BERT, which is fine-tuned by the human-labeled relevance data.
    \vspace*{-3mm}
    \item \textbf{Student model design.} We design a novel student model BERM corresponding to the score function $G(\cdot)$ in Def.~\ref{definition4}. Specifically, macro and micro matching embeddings are derived in BERM to capture the sentence-level and word-level relevance matching signal, respectively. 
    Also, based on the metapaths `$Q$-$I$-$Q$'' and `$I$-$Q$-$I$'', we design a node-level encoder and a metapath-instance-level aggregator to derive metapath embeddings.
    \vspace*{-3mm}
    \item \textbf{Online application.} To serve online search, we conduct further distillation to BERM and obtain BERM-O, which can be easily deployed to the online search platform.
\end{itemize}

\subsection{BERM Model}
\label{BERM Model}
Next we describe BERM in detail, including 0-order relevance modeling (Section~\ref{0-order Relevance Modeling}), $k$-order relevance modeling (Section~\ref{k-order Relevance Modeling}), and overall learning objective (Section~\ref{The Final Learning Objective}).

\subsubsection{0-order Relevance Modeling}
\label{0-order Relevance Modeling}
The whole structure of BERM includes both the 0-order relevance modeling and $k$-order relevance modeling. This subsection introduces the 0-order relevance modeling which captures sentence-level and word-level matching signals by incorporating the macro matching embedding and micro matching embedding, respectively.

\textbf{Word embedding in e-commerce scene.} 
In e-commerce scene, the basic representations of query or item is an intractable problem.
On one hand, it is infeasible to represent queries and items as individual embeddings due to the unbounded entity space. On the other hand, product type names (like ``iphone11'') or attribute names (like ``256GB'') have special background information and could contain complex lexicons such as different languages and numerals.
To address these problems, we adopt word embedding in BERM, which dramatically reduces the representation space. Also, we treat contiguous numerals, contiguous English letters, or single Chinese characters as one word and only retain the high-frequency words (such as the words occurring more than fifty times in a six-month search log) in the vocabulary. The final vocabulary is only in the tens of thousands, which saves memory consumption and lookup time of indexes by a large margin. 
Each word is represented by a $d$-dimensional embedding vector, which is trained by Word2Vec~\cite{24}. The $i$-th word's embedding of query $Q$ (or item title $I$) is denoted as \scalebox{0.9}{$\bm{E}_{Q}^{i}\in \mathbb{R}^{d}$} (or \scalebox{0.9}{$\bm{E}_{I}^{i}\in \mathbb{R}^{d}$}).

\textbf{Macro and micro matching embeddings.} To capture sentence-level and word-level matching signals, we employ macro matching embedding and micro matching embedding, respectively. For the macro matching embedding, taking query $Q$ with $l_Q$ words and item $I$ with $l_I$ words as examples, their macro embeddings \scalebox{0.9}{${\boldsymbol{E}}_{\mathrm{seq}}^{Q},{\boldsymbol{E}}_{\mathrm{seq}}^{I} \in \mathbb{R}^{d}$} are calculated by the column-wise mean value of \scalebox{0.9}{${\boldsymbol{E}}_{Q} \in \mathbb{R}^{l_{Q}\times d},{\boldsymbol{E}_{I}} \in \mathbb{R}^{l_{I}\times d}$}:

\vspace{-3mm}
\begin{small}
\begin{equation}
\setlength{\abovedisplayskip}{1pt}
\setlength{\belowdisplayskip}{1pt}
\bm{E}_{\mathrm{seq}}^{Q}=\frac{1}{l_Q}\sum_{i=1}^{l_Q}\bm{E}_{Q}^{i},\quad
\bm{E}_{\mathrm{seq}}^{I}=\frac{1}{l_I}\sum_{i=1}^{l_I}\bm{E}_{I}^{i} \;.
\end{equation}
\end{small}For the micro matching embedding, we first build an interaction matrix \scalebox{0.9}{$\bm{M}_{\mathrm{int}} \in \mathbb{R}^{l_{Q}\times l_{I}}$} whose \scalebox{0.9}{$(i,j)$}-th entry is the dot product of \scalebox{0.9}{$\bm{E}_{Q}^{i}$ and $\bm{E}_{I}^{j}$}:

\vspace{-2mm}
\begin{small}
\begin{equation}
\setlength{\abovedisplayskip}{1pt}
\setlength{\belowdisplayskip}{1pt}
\bm{M}_{\mathrm{int}}=\left \{m_{\mathrm{int}}^{i,j}\right \}_{l_{Q}\times l_{I}},\quad
m_{\mathrm{int}}^{i,j}=\langle \bm{E}_{Q}^{i}, \bm{E}_{I}^{j}\rangle \;.
\end{equation}
\end{small}Then the micro matching embedding \scalebox{0.9}{$\bm{E}_{\mathrm{int}}\in \mathbb{R}^{l_{Q}l_{I}}$} is the vectorization of \scalebox{0.9}{$\bm{M}_{\mathrm{int}}$}, i.e., 
\scalebox{0.9}{$\bm{E}_{\mathrm{int}}=\mathrm{vec}(\bm{M}_{\mathrm{int}})$}.


\subsubsection{\textit{k}-order Relevance Modeling}
\label{k-order Relevance Modeling}
The $k$-order relevance model contains a node-level encoder and a metapath-instance-level aggregator.

\textbf{Node-level encoder.}
The input of the node-level encoder is node embeddings and its output is an instance embedding (i.e., the embedding of a metapath instance). Specifically, to obtain the instance embedding, we integrate the embeddings of neighboring nodes into the anchor node embedding with a mean encoder. Taking ``$Q$-$I_\mathrm{{top1}}$-$Q_\mathrm{{top1}}$'' (note that ``$I_\mathrm{{top1}}$'' is the top-1 node in the 1-hop neighbor list of node $Q$, and see Appendix~\ref{Bipartite Graph Construction} for further details) as an example, its embedding \scalebox{0.9}{$\bm{E}_{Q-I_\mathrm{{top1}}-Q_\mathrm{{top1}}}\in \mathbb{R}^{d}$} is calculated by:

\vspace{-2mm}
\begin{small}
\begin{equation}
\setlength{\abovedisplayskip}{2pt}
\setlength{\belowdisplayskip}{2pt}
\bm{E}_{Q-I_\mathrm{{top1}}-Q_\mathrm{{top1}}}=\mathrm{MEAN}(\bm{E}_{\mathrm{seq}}^{Q}, \bm{E}_{\mathrm{seq}}^{I_{\mathrm{top1}}},\bm{E}_{\mathrm{seq}}^{Q_{\mathrm{top1}}}).
\end{equation}
\end{small}The metapath instance bridges the communication gap between different types of nodes and can be used to update the anchor node embedding from structure information.

\begin{figure}[t]
\centering
\includegraphics[width=0.40\textwidth]{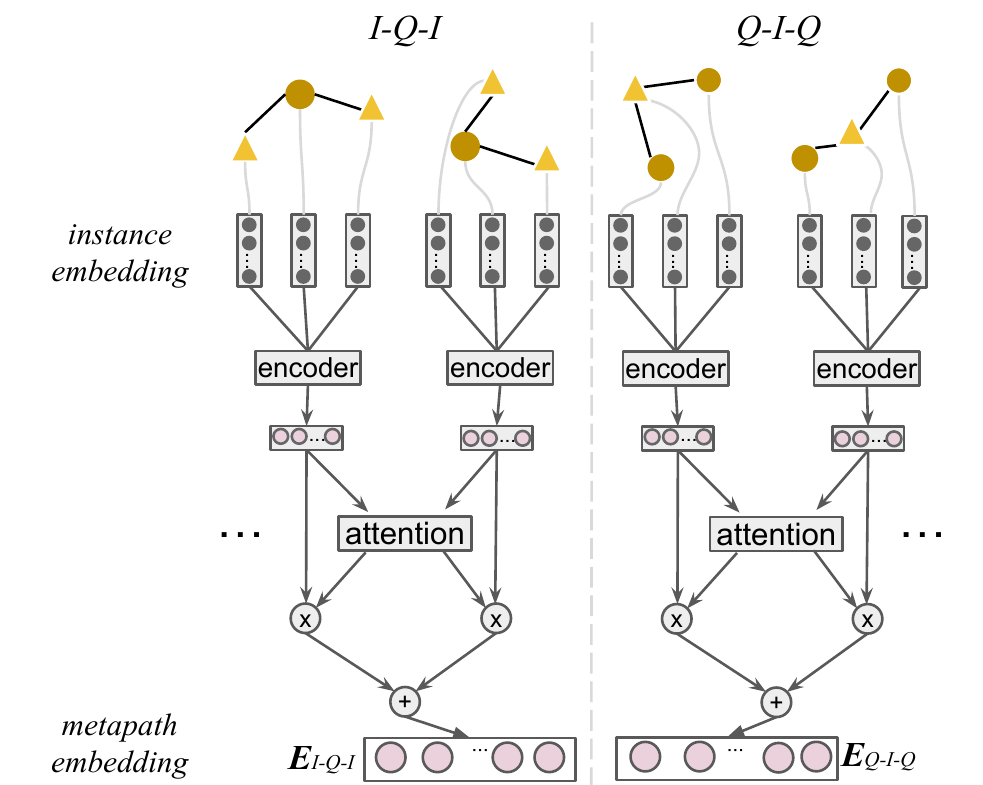}
\hspace{0in}
\caption{Calculation process of $\bm{E}_{Q-I-Q}$ and $\bm{E}_{I-Q-I}$ in BERM.}
\label{model3}
\vspace*{-4mm}
\end{figure}

\textbf{Metapath-instance-level aggregator.}
The inputs of the metapath-instance-level aggregator are instance embeddings and its output is a metapath embedding.
Different metapath instances convey different information, so they have various effects on the final metapath embedding. However, the mapping relationship between the instance embedding and metapath embedding is unknown. To learn their relationship automatically, we introduce the ``graph attention'' mechanism to generate metapath embeddings~\cite{attention_1,attention_2}.
Taking metapath ``$Q$-$I$-$Q$'' as an example, we use graph attention to represent the mapping relationship between ``$Q$-$I$-$Q$'' and its instances. The final metapath embedding \scalebox{0.9}{$\bm{E}_{Q-I-Q} \in \mathbb{R}^{d}$} is obtained (\scalebox{0.9}{$\bm E_{I-Q-I}\in \mathbb{R}^{d}$} is calculated similarly) by accumulating all instance embeddings with attention scores $\mathrm{Att}_{1},\mathrm{Att}_{2},\mathrm{Att}_{3},\mathrm{Att}_{4}\in \mathbb{R}$:

\vspace{-2mm}
\begin{small}
\begin{equation}
\setlength{\abovedisplayskip}{1pt}
\setlength{\belowdisplayskip}{1pt}
\begin{aligned}
\bm{E}_{Q-I-Q}&= \mathrm{Leaky ReLU}(\mathrm{Att}_{1}\cdot \bm{E}_{Q-I_\mathrm{{top1}}-Q_\mathrm{{top1}}} \\
&+\mathrm{Att}_{2}\cdot \bm{E}_{Q-I_\mathrm{{top1}}-Q_\mathrm{{top2}}} \\
&+\mathrm{Att}_{3}\cdot \bm{E}_{Q-I_\mathrm{{top2}}-Q_\mathrm{{top1}}} \\
&+\mathrm{Att}_{4}\cdot \bm{E}_{Q-I_\mathrm{{top2}}-Q_\mathrm{{top2}}}).
\end{aligned}
\end{equation}
\end{small}Though $\mathrm{Att}_{i}$ can be set as a fixed value, we adopt a more flexible way, i.e., using the neural network to learn $\mathrm{Att}_{i}$ automatically. Specifically, we feed the concatenation of the anchor node embedding and metapath instance embedding into a one-layer neural network (its weight is \scalebox{0.9}{$\bm{W}_{\mathrm{att}}\in \mathbb{R}^{6d\times 4}$} and its bias is \scalebox{0.9}{$\bm{b}_{\mathrm{att}}\in \mathbb{R}^{1\times 4}$}) with a softmax layer, which outputs an attention distribution:

\vspace{-2mm}
\begin{small}
\begin{equation}
\setlength{\abovedisplayskip}{1pt}
\setlength{\belowdisplayskip}{0.5pt}
(\mathrm{Att}_{i})_{1\leq i \leq 4}=\mathrm{softmax}(\bm{E}_\mathrm{concat}*\bm{W}_{\mathrm{att}}+\bm{b}_{\mathrm{att}}),
\end{equation}
\end{small}

\vspace{-2mm}
\begin{small}
\begin{equation}
\setlength{\abovedisplayskip}{0.5pt}
\setlength{\belowdisplayskip}{1pt}
\begin{aligned}
\bm{E}_\mathrm{concat}&=[\bm{E}_{\mathrm{seq}}^{Q}|\bm{E}_{\mathrm{seq}}^{I}|\bm{E}_{Q-I_\mathrm{{top1}}-Q_\mathrm{{top1}}}|\bm{E}_{Q-I_\mathrm{{top1}}-Q_\mathrm{{top2}}} \\
&| \bm{E}_{Q-I_\mathrm{{top2}}-Q_\mathrm{{top1}}}|\bm{E}_{Q-I_\mathrm{{top2}}-Q_\mathrm{{top2}}}].
\end{aligned}
\end{equation}
\end{small}The above process is shown in Fig.~\ref{model3}.

\textbf{Embedding fusion.}
By the 0-order and $k$-order relevance modeling, three types of embeddings are generated, including macro matching embedding (\scalebox{0.9}{$\bm{E}^{Q}_{\mathrm{seq}},\bm{E}^{I}_{\mathrm{seq}}\in \mathbb{R}^{d}$}), micro matching embedding (\scalebox{0.9}{$\bm{E}_{\mathrm{int}}\in \mathbb{R}^{l_{Q}l_{I}}$}), and metapath embedding (\scalebox{0.9}{$\bm{E}_{Q-I-Q},\bm{E}_{I-Q-I}\in \mathbb{R}^{d}$}). We concatenate them together and feed the result to a three-layer neural network (its weights are \scalebox{0.9}{$\bm{W}_{0}\in \mathbb{R}^{(4d+l_{Q}l_{I})\times d},\bm{W}_{1},\bm{W}_{2}\in \mathbb{R}^{d\times d},\bm{W}_{3}\in \mathbb{R}^{d\times 1}$} and biases are \scalebox{0.9}{$\bm{b}_{\mathrm{0}},\bm{b}_{\mathrm{1}},\bm{b}_{\mathrm{2}}\in \mathbb{R}^{1\times d},\bm{b}_{\mathrm{3}}\in \mathbb{R}^{1\times 1}$}), which outputs the final relevance estimation score $\hat{y}_{i}$:

\vspace{-2mm}
\begin{small}
\begin{equation}
\setlength{\abovedisplayskip}{1pt}
\setlength{\belowdisplayskip}{1pt}
    \hat{y}_{i} = \mathrm{Sigmoid}(\bm{E}_{3}*\bm W_{3}+\bm b_{3}),
    \label{relevance_score}
\end{equation}
\vspace{-3mm}
\begin{equation}
\setlength{\abovedisplayskip}{1pt}
\setlength{\belowdisplayskip}{1pt}
\hspace{-1mm}
    \bm{E}_{j+1} = \mathrm{ReLU}(\bm{E}_{j}*\bm W_{j}+\bm b_{j}),
    \bm{E}_{0} = \bm{E}_{all}, j=0,1,2,
    \label{FFN}
\end{equation}
\vspace{-2mm}
\begin{equation}
\setlength{\abovedisplayskip}{1pt}
\setlength{\belowdisplayskip}{1pt}
    \bm{E}_{all} = [\bm{E}^{Q}_{\mathrm{seq}}|\bm{E}^{I}_{\mathrm{seq}}|\bm{E}_{\mathrm{int}}|\bm{E}_{Q-I-Q}|\bm{E}_{I-Q-I}].
    \label{concat_all}
\end{equation}
\end{small}

\subsubsection{Overall Learning Objective}
\label{The Final Learning Objective}
We evaluate the cross-entropy error on the estimation score $\hat{y}_{i}$ and label $y_{i}$ (note that $y_{i}\in [0,1]$ is the score of the teacher model BERT, see Appendix~\ref{Bipartite Graph Construction} for further details), and then minimize the following loss function:

\vspace{-2mm}
\begin{small}
\begin{equation}
\setlength{\abovedisplayskip}{1pt}
\setlength{\belowdisplayskip}{1pt}
L = -\sum_{i=1}^{\tilde{n}} y_{i}\log(\hat{y}_{i})+(1-y_{i})\log(1-\hat{y}_{i}).
\end{equation}
\end{small}In Appendix~\ref{Complexity Analysis}, we analyze the complexities of BERT, BERM, and BERM-O.

\section{Experiments}
\label{experiments}
In this section, we present the offline and online experimental results of BERM~\footnote{We provide the description of baselines, implementation details, and additional experiments in Appendix~\ref{Baselines},~\ref{Implementation Details},~\ref{Additional Experiments} (Code URL: https://github.com/Young0222/EMNLP-BERM).}.

\subsection{Experimental Setting}
\textbf{Datasets.}
We collect three datasets from the anonymous platform, including the ``Electronics'' category (Data-E), all-category (Data-A), and sampled all-category (Data-S).
In the platform, there are mainly three different levels of item categories: $Cid_1$ (highest level, e.g., ``Electronics''), $Cid_2$ (e.g., ``Mobile phone''), and $Cid_3$ (lowest level, e.g., ``5G phone''). Data-A, Data-S, and Data-E have different data distributions.
Specifically, Data-A covers all first-level categories $Cid_1$ in the platform, Data-S is generated by uniformly sampling 5,000 items from $Cid_1$, and Data-E only focuses on ``Electronics'' in $Cid_1$.

For the training data $S_{\mathrm{train}}$ (also called $S_{\mathrm{transfer}}$), the collected user behaviors include click and purchase. For the testing data $S_{\mathrm{test}}$, whose queries are disjointed with those of $S_{\mathrm{train}}$, we use human labeling to distinguish between relevant and irrelevant items. Specifically, editors are asked to assess the relevance scores between queries and items.
In the anonymous platform, the candidate set of relevance scores is $\left \{1,2,3,4,5\right \}$, where $5$ means most relevant and $1$ means least relevant. To simplify it, we use binary labeling including the positive label (i.e., 4 or 5) and negative label (i.e., 1, 2, or 3). We report the detailed statistics of Data-E, Data-A, and Data-S in Tab.~\ref{table-data}.

\vspace{-0cm}
\begin{table}[t]
\setlength\tabcolsep{3.2pt}
\scriptsize
\caption{Statistics of the used datasets.}
\vspace*{-2mm}
\begin{center}
\begin{tabular}{rrrrr}
\toprule[1.0pt]
\small \bf Sets &\small \bf Names  &\small \bf Data-E &\small \bf Data-A &\small \bf Data-S \\ 
\midrule[0.5pt]
\multirow{6}*{$S_{\mathrm{train}}$} &\# Example &6,369,396 &174,863,375 &11,397,439 \\
&\# $\mathrm{Node_{query}}$	&398,824 &5,952,020 &3,284,480  \\
&\# $\mathrm{Node_{item}}$ 	&728,405 &49,517,217 &1,307,557 \\
&\# Edge 	&5,070,460 &159,205,320 &7,525,355 \\ 
&\# Click 	&1,471,079,596 &5,109,731,591 &1,431,899,847 \\
&\# Purchase 	&33,285,887 &322,151,488 &118,495,170 \\
\midrule[0.5pt]
\multirow{4}*{$S_{\mathrm{test}}$} &\# Example &30,563 &39,743 &39,743 \\ 
&\# $\mathrm{Node_{query}}$ 	&3,374 &3,108 &3,108 \\
&\# $\mathrm{Node_{item}}$ 	&16,137 &30,097 &30,097 \\
&\# Edge 	&18,988 &30,661 &30,661 \\
\bottomrule[1.0pt]
\end{tabular}
\end{center}
\label{table-data}
\vspace*{-5mm}
\end{table}

\textbf{Evaluation Metrics.}
To measure the performance of baseline methods and our BERM, we use three kinds of evaluation metrics, including Area Under the receiver operating characteristic Curve (AUC), F1-score, and False Negative Rate (FNR). 
The low value of FNR indicates the low probability of fetching irrelevant items, which is closely related to the user's search experience. Therefore, we include it in the evaluation metrics.

\begin{table}[t]
\setlength\tabcolsep{1.0pt}
\scriptsize
\caption{Comparisons on Data-E and Data-S. In each column, the best result is bolded and the runner-up is underlined. The symbol of ``$\downarrow$'' represents that the lower value corresponds to better performance. ``I, II, III'' represent the representation-focused, interaction-focused, and both-focused relevance matching models, respectively. ``IV'' represents the graph neural network models.}
\vspace*{-2mm}
\begin{center}
\begin{tabular}{cccccccc}
\toprule[1.0pt]
&\multirow{2}*{\small \bf Models}   &\multicolumn{3}{c}{\small \bf Data-E} &\multicolumn{3}{c}{\small \bf Data-S}\\
&&\bf AUC &\bf F1-score &\bf FNR($\downarrow$) &\bf AUC &\bf F1-score &\bf FNR($\downarrow$)\\
\midrule[0.5pt]
\multirow{3}*{I} &DSSM &0.6246 &0.6923 &0.9953 &0.8219 &0.8691 &1.0000 \\
&MVLSTM &\underline{0.8602} &\underline{0.8055} &0.3416 &0.7877 &0.8857 &0.7802 \\
&ARC-I	&0.8343 &0.7949 &0.3857 &0.6919	&0.8750 &0.9388 \\
\midrule[0.5pt]
\multirow{7}*{II} &DRMM  &0.6720 &0.6891 &0.7692 &0.6781 &0.8722 &0.9401 \\
&MatchPyramid	&0.7826 &0.7481 &0.5615 &0.7859 &0.8786 &0.8475\\
&ARC-II	&0.8128 &0.7864 &0.4377 &0.7606 &0.8784 &0.9076\\
&K-NRM	&0.7462 &0.7291 &0.6510 &0.7314 &0.8733 &0.9081 \\
&DRMM-TKS	&0.7678 &0.7383 &0.5462 &0.7793 &0.8789 &0.7893 \\
&Conv-KNRM &0.8369 &0.7879 &0.3469 &0.8029 &0.8789 &0.8913 \\
&ESIM &0.8056 &0.7769 &\underline{0.3373} &0.7987 &0.8623 &1.0000 \\
\midrule[0.5pt]
\multirow{2}*{III} &Duet &0.7693 &0.7219 &0.8173 &0.7968 &0.8754 &0.9458 \\
&BERT2DNN &0.8595 &0.8037 &0.3464 &\underline{0.8313} &\underline{0.9061} &\underline{0.4450} \\
\midrule[0.5pt]
\multirow{5}*{IV} &GAT &0.7526 &0.7361 &0.7529 &0.7411 &0.8746 &0.9234 \\
&GraphSAGE-Mean &0.7493 &0.7330 &0.7422 &0.7406 &0.8719 &0.9119 \\
&GraphSAGE-LSTM &0.7588 &0.7509 &0.6536 &0.7529 &0.8743 &0.8652 \\
&TextGNN &0.8310 &0.8029 &0.4525 &0.8277 &0.8779 &0.7549 \\
&GEPS &0.8405 &0.8037 &0.4892 &0.8254 &0.8794 &0.6340 \\
\midrule[0.5pt]
& BERM &\textbf{0.8785} &\textbf{0.8256} &\textbf{0.2966} &\textbf{0.8758} &\textbf{0.9079} &\textbf{0.3625} \\
\bottomrule[1.0pt]
\end{tabular}
\end{center}
\label{table-compare1}
\vspace*{-5mm}
\end{table}

\begin{table*}[t]
\setlength\tabcolsep{5pt}
\scriptsize
\setlength{\abovecaptionskip}{-0cm} 
\setlength{\belowcaptionskip}{-0mm}
\caption{Cases of e-commerce product search. ``$y_{i}$'' is the prediction score of the teacher model BERT and ``$\hat{y}_{i}$'' is the relevance estimation score of the student model BERM.}
\begin{center}
\begin{tabular}{ccccc}
\toprule[1.0pt]
\small \bf Query &\small \bf Item title &\small \bf Human labeling &\small \bf $y_{i}$ &\small \bf $\hat{y}_{i}$ \\
\bottomrule[0.5pt]
whistle &Li Ning whistle for basketball or volleyball game &Positive &0.9961 &0.9681 \\
women's dance shoe & Sansha modern dance shoe P22LS (black, women) &Positive &0.9973 &0.9908 \\
violin adult & FineLegend 1/8 violin FLV1114 &Positive &0.9950 &0.9769 \\
skating knee panels &RMT sports knee panels (black, L size) &Positive &0.9652 &0.9841 \\
DJI g1200a &DJI Mavic Mini unmanned aerial vehicle &Negative &0.0049 &0.0514 \\
Berkshire Hathaway Letters to Shareholders &The Snowball: Warren Buffett and the Business of Life &Negative &0.0258 &0.0874 \\
My brother called Shun Liu, ZHU SU JIN &Brothers: A Novel; Author: Yu Hua &Negative &0.1624 &0.0263 \\
nissan thermos cup &Disney thermos cup 500ML &Negative &0.4938 &0.2513 \\
java web exercises &JSP project development Case Full Record &Negative &0.5106 &0.3111 \\
\bottomrule[1.0pt]
\end{tabular}
\end{center}
\label{case-study}
\vspace*{-2mm}
\end{table*}

\subsection{Offline Performance}
\label{offline-performance}
We compare BERM with 12 state-of-the-art relevance matching methods and 5 graph neural network models on our in-house product search data. The results are shown in Tab.~\ref{table-compare1}. Because some baseline methods (e.g., DRMM and ESIM) have high time complexities, we use Data-E and Data-S for training and testing models.

As shown in Tab.~\ref{table-compare1}, BERM outperforms all the baselines according to the metrics of AUC, F1-score, and FNR. More specifically, we have the following findings: 1) Compared to the second-best method MVLSTM (BERT2DNN), BERM surpasses it 1.83\% (4.45\%) according to AUC on Data-E (Data-S). Furthermore, BERM achieves the lowest value of FNR on both Data-E and Data-S. This implies that BERM can easily identify irrelevant items so that it can return a list of satisfactory items in the real-world scene. 2) The collected training data have imbalanced classes (i.e., the positive examples are far more than the negative examples), which poses a challenge to model learning. Most baselines are sensitive to class imbalance. Since BERM learns explicit node semantics by integrating the neighboring node information, our method is robust when the data are imbalanced.

\vspace*{-1mm}
\subsection{Case Study}
\label{Case Study}
Apart from the above quantitative analysis, we conduct qualitative analysis based on some cases of e-commerce product search. For these cases, we list the query phrase, item title, human labeling, score of BERT, and score of BERM in Tab.~\ref{case-study}. We have the following empirical conclusions: 1) Most of the student's scores are close to the teacher's, which indicates the success of the proposed knowledge distillation framework. 2) Some cases imply that context information is necessary for relevance matching. For example, for the query ``nissan thermos cup'', the teacher model cannot explicitly judge whether or not the item entitled ``Disney thermos cup 500ML'' is relevant to it. With the help of context information in the query-item bipartite graph, BERM can recognize that this query is related to ``nissan'', rather than ``Disney''.

\begin{table}[t]
\setlength\tabcolsep{4pt}
\scriptsize
\setlength{\abovecaptionskip}{-0cm} 
\setlength{\belowcaptionskip}{-0mm}
\caption{Online performance of BERM-O under price sort mode and default sort mode.}
\begin{center}
\begin{tabular}{ccccc}
\toprule[1.0pt]
\multirow{2}*{\small \bf Metrics} &\multicolumn{2}{c}{\small \bf Price Sort Mode} &\multicolumn{2}{c}{\small \bf Default Sort Mode}\\
 &\bf Improvement &\bf P-value &\bf Improvement &\bf P-value\\
\midrule[0.5pt]
UV-value &5.713\% &3.20e-2 &0.5013\%&1.10e-1\\
UCVR &1.540\% &7.81e-2 &0.3058\%&1.75e-2\\
CVR &1.829\% &1.01e-2 &0.1218\%&1.60e-1\\
RPM &5.587\% &3.03e-2 &0.6886\%&2.32e-2\\
\bottomrule[1.0pt]
\end{tabular}
\end{center}
\label{online}
\vspace*{-2mm}
\end{table}

\subsection{Deployment \& Online A/B Testing}
\label{Deployment & Online A/B Testing}
We conduct further distillation to BERM and obtain a lighter model BERM-O whose basic structure is a two-layer neural network. The process of further distillation is almost the same as the first knowledge distillation. The transfer set generated by further distillation has graph-context labels.
To further evaluate BERM-O's performance in the real search scene, we deploy it to the anonymous online search platform.
On this platform, there are about one hundred million daily active users (DAU) and two billion items. It processes over 150 million search queries per day. The online baseline group BERT2DNN~\cite{bert2dnn} and control group BERM-O are deployed in a cluster, where each node is with 64 core Intel(R) Xeon(R) CPU E5-2683 v4 @ 2.10GHz, 256GB RAM as well as 4 NVIDIA TESLA P40 GPU cards. For both groups, the only needed input data are queries and item titles, which can be easily caught from the online environment. Since BERM-O is lighter than BERT or BERM, deploying it to the online search chain requires less engineering work in the system.

\textbf{Online results.} We compare BERM-O with BERT2DNN~\cite{bert2dnn} which is our online baseline model using knowledge distillation without context information. The results of A/B testing are reported in Tab.~\ref{online}. These results are from one observation lasting more than ten days.
Four widely-used online business metrics are adopted 1) conversion rate (CVR): the average order number of each click behavior, 2) user conversion rate (UCVR): the average order number of each user, 3) unique visitor value (UV-value): the average gross merchandise volume of each user, and 4) revenue per mile (RPM): the average gross merchandise volume of each retrieval behavior.
The results show that BERM-O outperforms BERT2DNN in the platform according to all of the business metrics.
For example, BERM-O significantly improves 5.7\% (relative value) of UV-value under price sort mode.

\vspace{4mm}
\section{Conclusions and Future Work}
\label{Conclusions and Future Work}
In this paper, we propose the new problem of contextual relevance matching in e-commerce product search.
Different from the previous work only using the 0-order relevance modeling, we propose a novel method of the $k$-order relevance modeling, i.e., employing bipartite graph embedding to exploit the potential context information in the query-item bipartite graph.
Compared to the state-of-the-art relevance matching methods, the new method BERM performs robustly in the experiments.
We deploy BERM-O (distilled from BERM) to the anonymous online e-commerce product search platform. The results of A/B testing indicate that BERM-O improves the user's search experience significantly.
In the future, we plan to apply our method to other e-commerce applications such as recommendation systems and advertisements.


\newpage
\appendix

\section{Related Work}
\label{Related Work}
In this section, we review classical relevance matching models (Section~\ref{Classical Relevance Matching Models}), Transformer-style models (Section~\ref{Transformer-style Models}), and online knowledge distillation methods (Section~\ref{Online Knowledge Distillation Methods}). In each subsection, we conclude the difference between the previous work and ours.
\subsection{Classical Relevance Matching Models}
\label{Classical Relevance Matching Models}
Traditional relevance matching uses TF-IDF or BM25 to calculate the text similarity between the query and document. This type of method is mainly based on the word frequency in the text and cannot measure the similarity relationship between different words or phrases. With the development of deep learning, relevance matching combined with the characterization's advantage of vector representations on the semantic relationship is booming. The prevailing methods are either representation-focused (e.g., DSSM~\cite{dssm}, CDSSM~\cite{cdssm} and ARC-I~\cite{arc-i}) or interaction-focused (e.g., MatchPyramid~\cite{MatchPyramid}, ARC-II~\cite{arc-i} and ESIM~\cite{esim}). The representation-focused methods learn the low-dimension representations of both sentences and then predict their relationship by calculating the similarity (such as cosine similarity) of representations. The interaction-focused methods directly learn an interaction representation of both sentences based on the interaction calculation from word-level to sentence-level. These methods are typically effective for context-free relevance matching. 

However, contextual relevance matching is necessary to real applications such as e-commerce product search. The above methods ignore the inherent context information contained in search logs~\cite{search_log_tkde_1,search_log_1}. In this work, we incorporate the advantages of representation-based and interaction-based embeddings into BERM, which is focused on contextual relevance matching.

\subsection{Transformer-style Models} 
\label{Transformer-style Models} 
More recently, Transformer-based models~\cite{emnlp-t-1,emnlp-t-2,emnlp-t-3,emnlp-t-4} have achieved breakthroughs on many NLP tasks and reached human-level accuracy. The representative models include BERT~\cite{bert}, ERNIE~\cite{ernie}, and RoBERTa~\cite{roberta}. However, the multi-layer stacked Transformer structure in these models leads to high time complexity, so they are hard to deploy online. To effectively apply Transformer-based models, in this work, we use BERT to generate the labels of transfer set~\cite{distillation-hinton} as the supervised information for BERM; 2) refine the noisy behavior data to ensure that the constructed bipartite graph is credible. 
Additionally, GRMM~\cite{GRMM} and GHRM~\cite{GHRM} use graph information to enforce the relevance matching model for information retrieval.

However, the above work is essentially different from ours on the definition of graphs. In their constructed graph, nodes are unique words and edges are the co-occurrent relationships. In this work, we leverage query phrases (or item titles) as nodes and user behaviors as edges, which is more suitable for the product search problem.

\subsection{Online Knowledge Distillation Methods}
\label{Online Knowledge Distillation Methods}
Knowledge distillation is firstly proposed in~\cite{distillation-hinton}. Its main idea is to transfer the knowledge generated by a massive teacher model into a light student model. Because of the low complexity of the student model, it is easy to deploy the student model to the online platform. Considering the strong semantic understanding ability of BERT, some studies exploit the potential of BERT as the teacher model of knowledge distillation. Two types of design principles are general: isomorphic principle and isomeric principle. Specifically, the distillation methods that follow the isomorphic principle use the same model architecture for teacher and student models, such as TinyBERT~\cite{tinybert}, BERT-PKD~\cite{BERT-PKD}, MTDNN~\cite{MTDNN}, and DistilBERT~\cite{distilbert}. As a more advanced design principle, the isomeric principle uses different model architectures for teacher and student models, such as Distilled BiLSTM~\cite{distilled-bilstm} and BERT2DNN~\cite{bert2dnn}.

Although the above methods reduce the total time costs by learning a light student model, they ignore the context information in either the teacher model or student model. Our proposed knowledge distillation framework follows the isomeric principle and further integrates the context information into the student model by bipartite graph embedding.


\section{Details of our Framework}
\label{Details of the framework}
In this section, we introduce more details about the proposed knowledge distillation framework, including the overview (Section~\ref{overview}), bipartite graph construction (Section~\ref{Bipartite Graph Construction}), and bipartite graph analysis (Section~\ref{Bipartite Graph Analysis}).

\subsection{Bipartite Graph Construction}
\label{Bipartite Graph Construction}
As we referred to in the body of the paper, queries and items in the product search can be naturally abstracted as a bipartite graph where edges are generated by multi-typed user behaviors like view, click, and purchase. However, user behaviors are biased and noisy. Directly learning the embedding on the raw user behavior graph $\mathcal{G}$ makes it difficult to estimate the relevance relationship between nodes. 
Many click models (see the survey in~\cite{survey-click}) have been proposed to absorb all the potential biases into their designed model. However, these click models are sophisticated, making them consume lots of memory resources in the online environment. To address this problem, we adopt a simple but effective strategy: we introduce the external knowledge from BERT to refine the raw user behavior graph $\mathcal{G}$ and then construct a knowledge-enhanced bipartite graph $\mathcal{G}'$. The whole graph construction includes the following phases.

\textbf{Fine-tuning BERT.}
Here the external knowledge is provided by the BERT model. Transformer-based models such as BERT~\cite{bert} and ERNIE~\cite{ernie} have been preferably used as NLP benchmarks in recent years. Here we use the BERT model as the teacher model, which is pre-trained on a large text corpus and then fine-tuned on our in-house data. The positive and negative examples in the data are human-labeled and cover various categories of items. The fine-tuned BERT is equipped with a great ability of relevance discrimination and thus can act as an expert on filtering noisy data. For each example pair $p_i$ in the transfer set $S_{\mathrm{transfer}}$, we use BERT to predict its score $y_i$ as the training label of the student model BERM.

\textbf{Behavior graph construction.}
The user behavior graph $\mathcal{G}$ is built on the user search log over six months which records click and purchase behaviors as well as their frequencies. In $\mathcal{G}$, an edge represents an existing click behavior or purchase behavior between a query and an item. The click behavior edges are dense and highly noisy, so we introduce the fine-tuned BERT model to refine $\mathcal{G}$.

\textbf{Knowledge-enhanced graph refinement.}
We retain all the raw purchase behavior edges, and meanwhile use the knowledge generated by the fine-tuned BERT to refine the click behavior edges. Specifically, we set two different thresholds $\alpha$ and $\beta$ to decide which click edges are removed and which important edges are added. This strategy can help remove the noise in user behaviors, and at the same time retrieve the missing but relevant neighbors which are not captured by user behaviors. To preserve the high-quality neighbor set, for each anchor node, we rank its 1-hop neighbors with the priority of ``purchase$>$high click$>$low click'' and sample the top two of them as the final neighbor list, e.g., the neighbor list of a query node $Q$ is represented as $[I_{\mathrm{top1}}, I_{\mathrm{top2}}]$. The implementation of graph construction is shown in Algorithm 1.

\begin{algorithm}[t]
\algsetup{linenosize=\scriptsize}
\small
\caption{Bipartite Graph Construction}
\begin{algorithmic}[1] 
\REQUIRE 
    Thresholds $\alpha$ and $\beta$; Collected dataset $S_{\mathrm{input}}=\left \{p_{i} \right \}_{\tilde{n}}$ (note that $\tilde{n}$ is the number of examples); Raw user behavior graph $\mathcal{G}=(\mathcal{U},\mathcal{V},\mathcal{E},\mathcal{A},\mathcal{R})$ where $\mathcal{E}=\left \{e_{1},\cdots,e_{m}\right \}$, $\mathcal{A}=\left \{Query, Item\right \}$, $\mathcal{R}=\left \{Click, Purchase\right \}$. \\
\ENSURE
    Transfer set $S_{\mathrm{transfer}}=\left \{ p_{i};y_{i} \right \}_{\tilde{n}}$ where $y_{i}$ is the training label of query-item pair $p_{i}$ ($y_{i}\in \left [ 0,1\right ]$); Refined bipartite graph ${\mathcal{G}}'=(\mathcal{U},\mathcal{V},{\mathcal{E}}',\mathcal{A},\mathcal{R})$. \\
\STATE Initialize ${\mathcal{E}}'$: ${\mathcal{E}}'={\mathcal{E}}$.  \\
\STATE Fine-tune BERT on the human-labeled data. \\
\STATE Use the fine-tuned BERT to predict on $S_{\mathrm{input}}$ and then obtain $S_{\mathrm{transfer}}=\left \{ p_{i};y_{i} \right \}_{\tilde{n}}$. \\
\FOR{$p_{i}, y_{i}$ in $S_{\mathrm{transfer}}$}
    \STATE Build an edge between the pair $p_{i}$ and denote it as $e_{i}=edge(p_{i})$.
    \IF {$e_{i} \in \mathcal{E}$}
        \IF {$f_{2}(e_{i})=Purchase$} \STATE continue;
        \ELSIF {$y_{i}<\alpha$} \STATE ${\mathcal{E}}'={\mathcal{E}}' \setminus \left \{{e_{i}}\right \}$
        \ENDIF
    \ELSIF {$y_{i}>\beta$}
        \STATE ${\mathcal{E}}'={\mathcal{E}}'\cup \left \{{e_{i}}\right \}$
    \ENDIF
\ENDFOR
\end{algorithmic}
\end{algorithm}

\subsection{Bipartite Graph Analysis}
\label{Bipartite Graph Analysis}
Network schema is an abstraction of the basic units in the graph~\cite{network_schema_ijcai}. In BERM, the basic network schema includes query node, item node, and the refined user-behavior-oriented edge between them. Based on this schema, we employ two metapaths: ``$Q$-$I$-$Q$'' and ``$I$-$Q$-$I$'' where ``$Q$'' and ``$I$'' represent ``$Query$'' and ``$Item$'', respectively. These two metapaths decide the used context information $C_{k}$ (Def.~\ref{definition4}) in our model. Compared to some more complex metapaths (such as ``$Q$-$I$-$Q$-$I$-$Q$'' and ``$I$-$Q$-$I$-$Q$-$I$''), the adopted metapaths in our model are both effective and computationally efficient. We further choose four instances for each metapath if they are available and pad them with zero embeddings if not. Taking ``$Q$-$I$-$Q$'' as an example, its instances include ``$Q$-$I_{\mathrm{top1}}$-$Q_\mathrm{{top1}}$'', ``$Q$-$I_\mathrm{{top1}}$-$Q_\mathrm{{top2}}$'', ``$Q$-$I_{\mathrm{top2}}$-$Q_\mathrm{{top1}}$'', and ``$Q$-$I_\mathrm{{top2}}$-$Q_\mathrm{{top2}}$''. For the incomplete metapaths, we pad them with null nodes which are not equipped with real semantic information.

\section{Complexity Analysis}
\label{Complexity Analysis}
In this section, we analyze the time and space complexities of BERT (teacher model), BERM (student model), and BERM-O (online model).

\subsection{Time Complexity}
\label{Time complexity}
For the lookup operation on the static vocabulary table (i.e., a word embedding table whose size is $n_w$), the time complexities of BERT, BERM, and BERM-O are the same, i.e., O($n_w$).
For the model calculation part, BERT uses Transformer networks.
We denote the word embedding size, head number, network number, query length, and item length as $d_{1}, h_{1}, k_{1}, l_{Q}, l_{I}$, respectively. For the one-layer multi-head attention mechanism, the complexity of linear mapping (input part) is O($\frac{1}{h_{1}}(l_{Q}+l_{I})d_{1}^{2}$), the complexity of attention operation is O($h_{1}l_{Q}^{2}d+hl_{I}^{2}d_{1}$), and the complexity of linear mapping (output part) is O($(l_{Q}+l_{I})d_{1}^{2}$). Therefore, the total model calculation complexity of $k_{1}$-layer BERT is \scalebox{0.9}{O($\frac{h_{1}+1}{h_{1}}(l_{Q}+l_{I})d_{1}^{2}k_{1} + (l_Q^2+l_I^2)h_{1}d_{1}k_{1}$)}.
For the student model BERM, we denote the word embedding size, hidden size, and network number as $d, h_{2}, k_{2}$, respectively.
The complexity of calculating micro matching embedding is O($l_{Q}l_{I}d$), which is far more than that of calculating micro matching embedding. The complexity of $k_{2}$-layer DNN is O($h_{2}dk_{2}$). Therefore, the total model calculation complexity of $k_{2}$-layer BERM is O($l_{Q}l_{I}d+h_{2}dk_{2}$).
For the online model BERM-O, we denote the word embedding size, hidden size, and network number as $d_{3}, h_{3}, k_{3}$, respectively.
The model calculation complexity of $k_{3}$-layer BERM-O is O($d_{3}h_{3}k_{3}$). Note that the complexity of BERM-O is independent of $l_{Q}$ and $l_{I}$ because BERM-O only receives sentence embeddings and does not calculate word-level matching signals.
Based on the above analysis, we can conclude that BERM is more efficient than BERT and meanwhile BERM-O has more advantages than BERM on time complexity.

\subsection{Space Complexity}
\label{Space complexity}
The storage of the static vocabulary table takes up the majority of the total space storage. Therefore, the space complexities of BERT, BERM, and BERM-O are the same, i.e., O($n_{w}d$).

\section{Details of Experimental setups}
\label{Details of Experimental setups}

\subsection{Baselines}
\label{Baselines}
The model BERM is compared with some state-of-the-art models. Like BERM, these models are used as the student model of the proposed knowledge distillation framework.
We adopt the hyper-parameter settings recommended by the original papers for all the methods. According to the formulation process of embedding, these methods can be divided into the following three types:
\begin{itemize}[leftmargin=*]
    \item Three representation-focused relevance matching methods (denoted by ``\textsuperscript{\S}'' in Tab.~\ref{table-compare1}): DSSM~\cite{dssm}, MVLSTM~\cite{MV-LSTM}, and ARC-I~\cite{arc-i}. They learn the low-dimension representations of both sentences w.r.t. a query and an item, and then predict their relationship by calculating the similarity (such as cosine similarity) of representations.
    \item Seven interaction-focused relevance matching methods (denoted by ``\textsuperscript{\dag}'' in Tab.~\ref{table-compare1}): DRMM~\cite{drmm}, MatchPyramid~\cite{match_pyramid}, ARC-II~\cite{ARC-I-ARC-II}, K-NRM~\cite{KNRM}, DRMM-TKS~\cite{drmm}, Conv-KNRM~\cite{KNRM}, and ESIM~\cite{esim}. They learn an interaction representation of both sentences based on the interaction calculation from word-level to sentence-level.
    \item Two integrated relevance matching methods (denoted by ``\textsuperscript{\ddag}'' in Tab.~\ref{table-compare1}): Duet~\cite{duet-model} and BERT2DNN~\cite{bert2dnn}. They combine the features of the above two types of methods into themselves.
    \item Five graph neural network models (denoted by ``\textsuperscript{$\star$}'' in Tab.~\ref{table-compare1}): GAT~\cite{gat}, GraphSAGE-Mean~\cite{graphsage}, GraphSAGE-LSTM~\cite{graphsage}, TextGNN~\cite{TextGNN}, and GEPS~\cite{GEPS}. They aggregate the neighbor information from the query graph or item graph to update the embedding of the anchor node.
\end{itemize}

\subsection{Implementation Details}
\label{Implementation Details}
Here we introduce the implementation details of the whole knowledge distillation framework as follows:
\begin{itemize}[leftmargin=*]
    \item \textbf{Teacher model}. For the teacher model, we adopt BERT-Base\footnote{https://github.com/google-research/bert} with a 12-layer ($k_{1}=12$) Transformer encoder where the word embedding size $d_{1}$ is 768 and head number $h_{1}$ is 12. We pre-train BERT-Base on a human-labeled dataset with 380,000 query-item pairs. The fine-tuned BERT-Base is then used as an expert to refine the noisy click behavior data from $S_\mathrm{train}$. The refinement rule is: if the prediction score $y_{i}$ of BERT-Base is less than $\alpha$ (the default value of $\alpha$ is 0.3), then the raw edge is deleted; if the score is larger than $\beta$ (the default value of $\beta$ is 0.7), then a new edge is added.
    \item \textbf{Student model}. For the student model, we adopt the proposed BERM model. We implement BERM in TensorFlow 2.0 with the high-level Estimator API. For each input query phrase $Q$ or item title $I$, we split it into several words and then truncate or pad its length to 10 or 65words (i.e., $l_{Q}=10, l_{I}=65$). Each word embedding is acquired by the lookup operation on a static vocabulary table whose total size $n_{w}$ is 39,846. This table is generated by pre-training two billion search data with the tool of Word2Vec. The size $d$ of pre-trained embeddings or trained embeddings is 128.
    
    \item \textbf{Training details}. We use Lazy-Adam as the optimizer and its learning rate is 0.001. To reduce the overfitting of the training data, we use L2 regularization on each layer of neural networks. For Data-E, we set the training epoch as 20. For Data-A and Data-S, we set the training epoch as 3. All results reported are selected according to the best AUC values on the testing set.
\end{itemize}

\section{Additional Experiments}
\label{Additional Experiments}
We conduct some additional experiments, including an ablation study (Section~\ref{Ablation Study}) and sensitivity analysis (Section~\ref{Sensitivity Analysis}). In these experiments, we adopt Data-E and Data-A consistently.

\subsection{Ablation Study}
\label{Ablation Study}

\subsubsection{Integration of Embeddings}
\label{Integration of embedding}
There are three types of components in the complete BERM: the representation-based embeddings $\bm{E}^{Q}_{\mathrm{seq}}, \bm{E}^{I}_{\mathrm{seq}}$, interaction-based embedding $\bm{E}_{\mathrm{int}}$, and metapath embeddings $\bm{E}_{Q-I-Q}, \bm{E}_{I-Q-I}$.
To further examine the importance of each component in the final embedding of BERM, we remove one or two components from it (Eq.~\ref{concat_all}) at a time and examine how the change affects its overall performance.

\begin{table}[t]
\setlength\tabcolsep{4pt}
\scriptsize
\setlength{\abovecaptionskip}{-0cm} 
\setlength{\belowcaptionskip}{-0mm}
\caption{Ablation study on Data-E. In each column, the best result is bolded.}
\begin{center}
\begin{tabular}{cccc}
\toprule[1.0pt]
\multirow{2}*{\small \bf Models}   &\multicolumn{3}{c}{\small \bf Data-E} \\
&\bf AUC &\bf F1-score &\bf FNR($\downarrow$)\\
\bottomrule[0.5pt]
$\bm{E}^{Q}_{\mathrm{seq}}, \bm{E}^{I}_{\mathrm{seq}}$	&0.8537	&0.8044	&0.3560 \\
$\bm{E}_{\mathrm{int}}$ &0.8595 &0.8037 &0.3464	\\
$\bm{E}_{Q-I-Q}, \bm{E}_{I-Q-I}$ &0.8430 &0.8173	&0.2995	\\
$\bm{E}^{Q}_{\mathrm{seq}}, \bm{E}^{I}_{\mathrm{seq}}, \bm{E}_{\mathrm{int}}$ &0.8638 &0.8086	&0.3331	\\
$\bm{E}_{\mathrm{int}}, \bm{E}_{Q-I-Q}, \bm{E}_{I-Q-I}$ &0.8761 &0.8221 &\textbf{0.2758} \\
$\bm{E}^{Q}_{\mathrm{seq}}, \bm{E}^{I}_{\mathrm{seq}}, \bm{E}_{Q-I-Q}, \bm{E}_{I-Q-I}$ &0.8656	&0.8190	&0.2922	\\
BERM &\textbf{0.8785} &\textbf{0.8256} &0.2966\\
\bottomrule[1.0pt]
\end{tabular}
\end{center}
\label{ablation-data-m}
\end{table}

\begin{table}[t]
\setlength\tabcolsep{4pt}
\scriptsize
\setlength{\abovecaptionskip}{-0cm} 
\setlength{\belowcaptionskip}{-0mm}
\caption{Ablation study on Data-A. In each column, the best result is bolded.}
\begin{center}
\begin{tabular}{cccc}
\toprule[1.0pt]
\multirow{2}*{\small \bf  Models} & \multicolumn{3}{c}{\small \bf  Data-A}   \\
&\bf AUC &\bf F1-score &\bf FNR($\downarrow$) \\
\bottomrule[0.5pt]
$\bm{E}^{Q}_{\mathrm{seq}}, \bm{E}^{I}_{\mathrm{seq}}$	&0.8067 &0.8660 &0.3697 \\
$\bm{E}_{\mathrm{int}}$ &0.8289 &0.9084 &0.4459 \\
$\bm{E}_{Q-I-Q}, \bm{E}_{I-Q-I}$ &0.8500	&\textbf{0.9114}	&0.4110 \\
$\bm{E}^{Q}_{\mathrm{seq}}, \bm{E}^{I}_{\mathrm{seq}}, \bm{E}_{\mathrm{int}}$ &0.8776	&0.9070	&0.4743 \\
$\bm{E}_{\mathrm{int}}, \bm{E}_{Q-I-Q}, \bm{E}_{I-Q-I}$ &0.8824	&0.9099	&0.3705 \\
$\bm{E}^{Q}_{\mathrm{seq}}, \bm{E}^{I}_{\mathrm{seq}}, \bm{E}_{Q-I-Q}, \bm{E}_{I-Q-I}$ &0.8750	&0.9094	&0.3753 \\
BERM &\textbf{0.8862} &0.9107 &\textbf{0.3673}\\
\bottomrule[1.0pt]
\end{tabular}
\end{center}
\label{ablation-data-a}
\end{table}

The corresponding results on Data-E and Data-A are reported in Tab.~\ref{ablation-data-m} and~\ref{ablation-data-a}. We have the following empirical observation and analysis:
\begin{itemize}[leftmargin=*]
\item In general, the both-component setting outperforms the single-component setting but is worse than the triple-component setting (i.e., BERM). It demonstrates that different components in BERM have different positive effects on the overall performance and they cannot replace each other.
\item The introduction of $k$-order relevance modeling can bring stable advancement to each 0-order relevance model. For example, the combination of ``\scalebox{0.9}{$\bm{E}^{Q}_{\mathrm{seq}}, \bm{E}^{I}_{\mathrm{seq}}, \bm{E}_{Q-I-Q}, \bm{E}_{I-Q-I}$}'' surpasses the combination of ``\scalebox{0.9}{$\bm{E}^{Q}_{\mathrm{seq}}, \bm{E}^{I}_{\mathrm{seq}}$}'' 6.83\% according to the metric of AUC on Data-A.
This demonstrates that applying metapath embedding to relevance matching can make effective use of the neighboring nodes' information in the user behavior graph.
\end{itemize}

\subsubsection{Effect of the Intermediate Node}
\label{Effect of the intermediate node}
The metapath defined in BERM includes the intermediate node. To further investigate the effect of the intermediate node, we compare the performances of BERM with the intermediate node (i.e., ``$Q$-$I$-$Q$'' and ``$I$-$Q$-$I$'') and BERM without the intermediate node (i.e., ``$Q$-$Q$'' and ``$I$-$I$'') on Data-E and Data-A in Fig.~\ref{intermediate}. We observe that BERM with the intermediate node performs better than the other one. We infer that the intermediate node has strong semantic closeness to the anchor node and thus it is helpful to accurate semantic recognition.

\begin{figure}[t]
\centering
\hspace{-5mm}
\subfigure[Data-E]{
\includegraphics[width=0.23\textwidth]{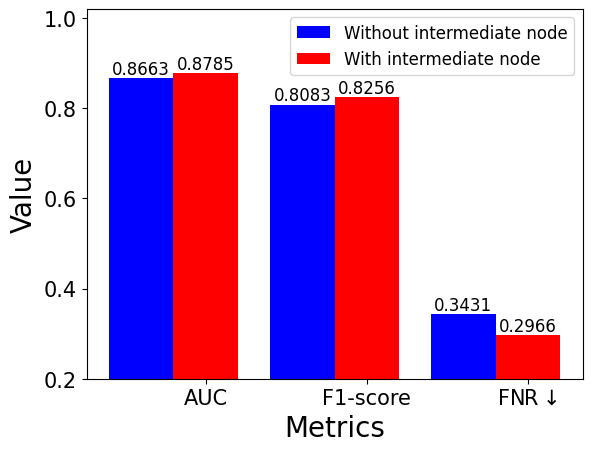}
\label{1}}
\hspace{-2mm}
\subfigure[Data-A]{
\includegraphics[width=0.23\textwidth]{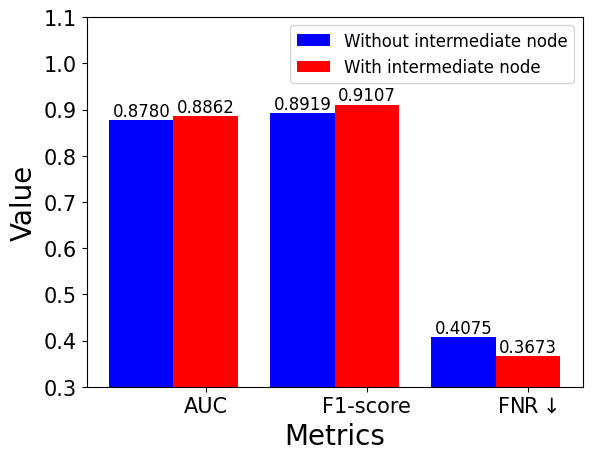}
\label{2}}
\hspace{-5mm}
\caption{Effect of the intermediate node. The red (blue) bar represents BERM with (without) the intermediate node.}
\label{intermediate}
\end{figure}

\subsection{Sensitivity Analysis}
\label{Sensitivity Analysis}
\subsubsection{Thresholds $\alpha$ and $\beta$}
In BERM, $\alpha$ decides how many edges of the noisy click behavior should be deleted and $\beta$ decides how many hidden useful edges should be retrieved. To investigate the sensitivity of $\alpha$ and $\beta$, we conduct experiments with 16 different hyper-parameter settings where $\alpha$ ranges from 0.2 to 0.5 and $\beta$ ranges from 0.5 to 0.8. We apply the three-order curve interpolation method to show the final results in Fig.~\ref{sensitivity-alpha}. In general, the results of BERM are robust to the change of hyper-parameter $\alpha$ and $\beta$ on either Data-E or Data-A. For example, the maximum error of AUC is no more than 1\%.
So we conclude that user behaviors play a major role in the performance of BERM and the knowledge from BERT provides auxiliary effects for it.

\begin{figure}[t]
\centering
\setlength{\abovecaptionskip}{0pt} 
\setlength{\belowcaptionskip}{0pt}

\subfigure[AUC of Data-E]{
\includegraphics[width=0.14\textwidth]{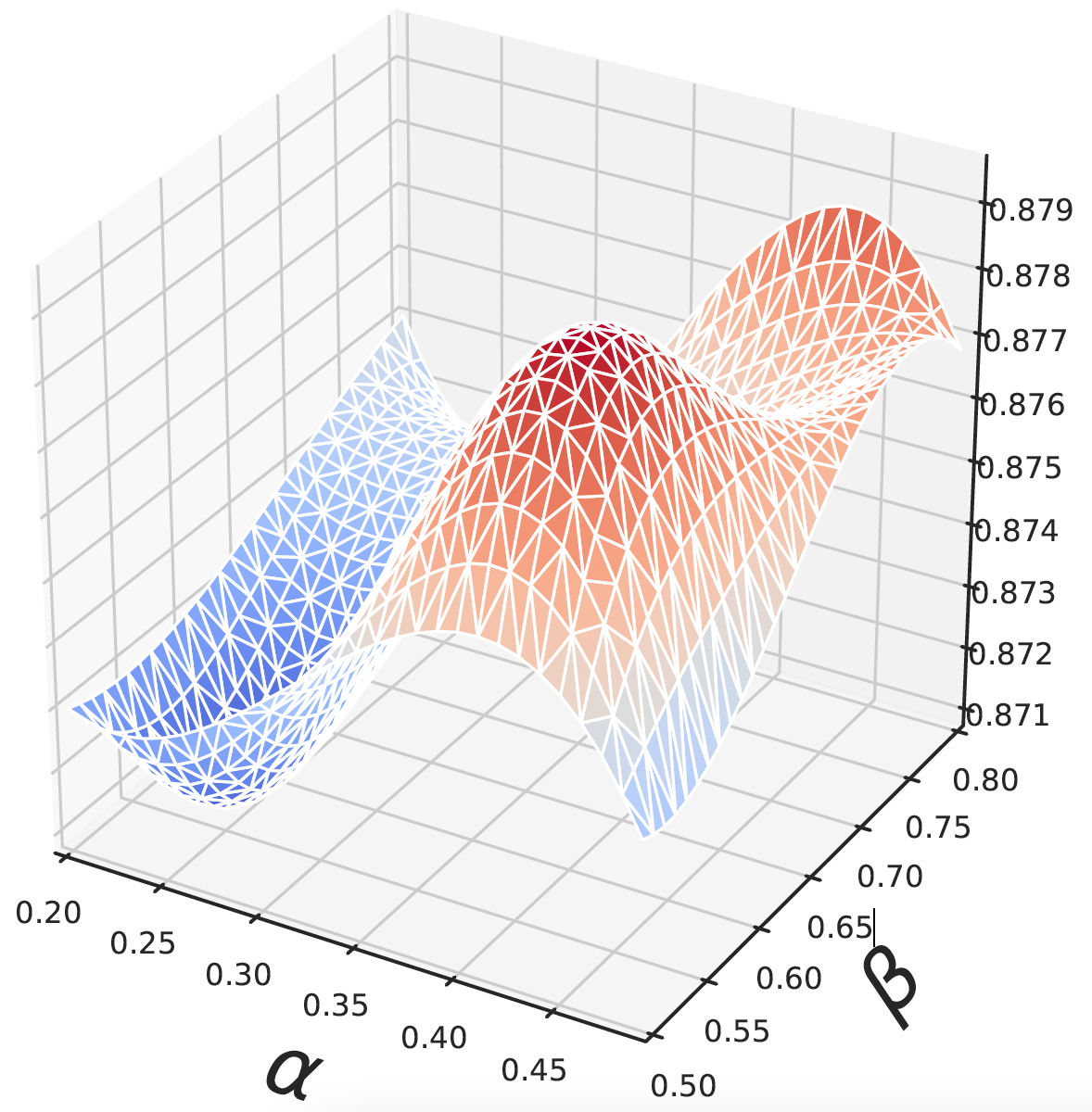}
}
\hspace{-1mm}
\subfigure[F1-score of Data-E]{
\includegraphics[width=0.14\textwidth]{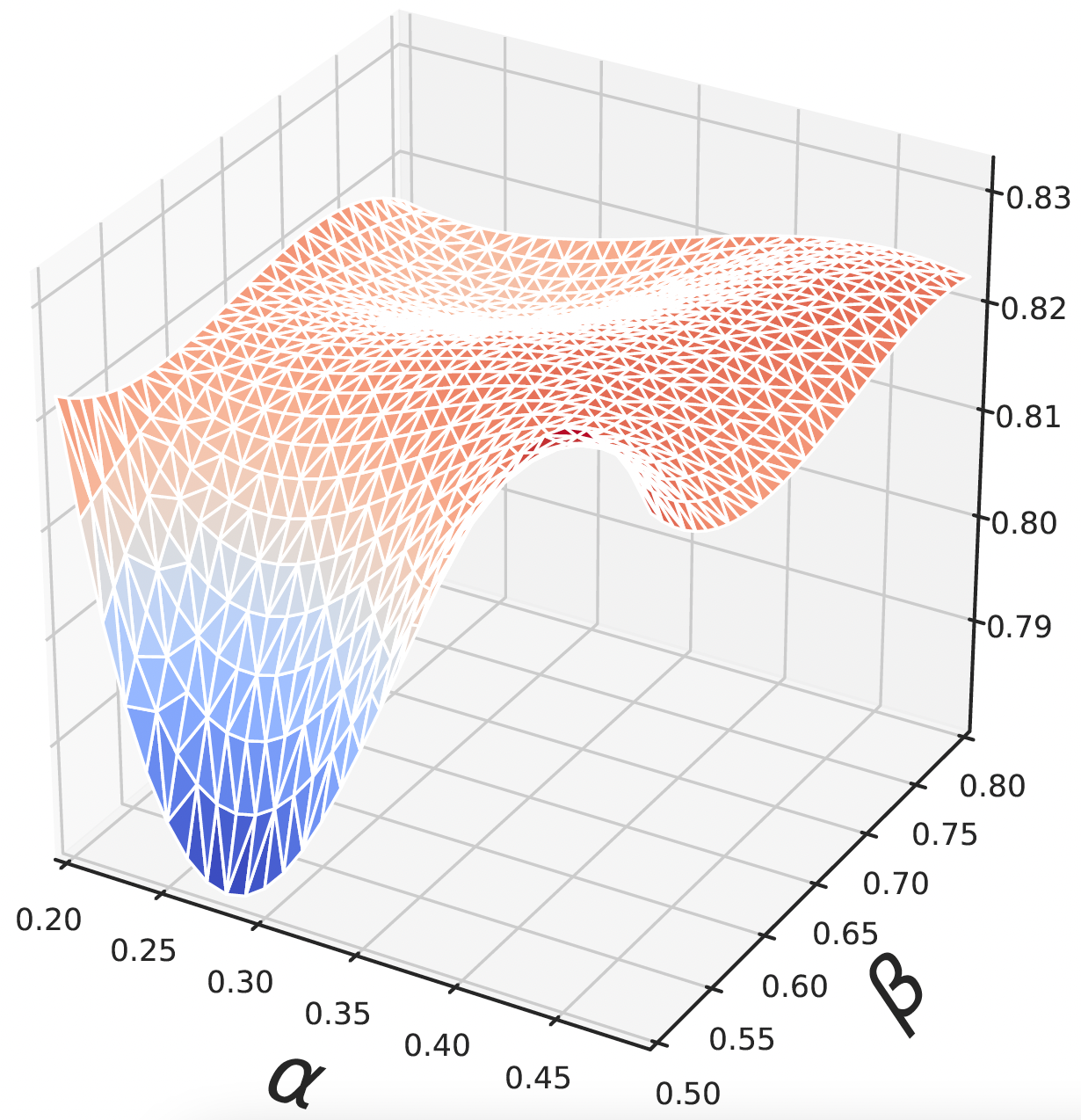}
}
\hspace{-1mm}
\subfigure[FNR of Data-E]{
\includegraphics[width=0.14\textwidth]{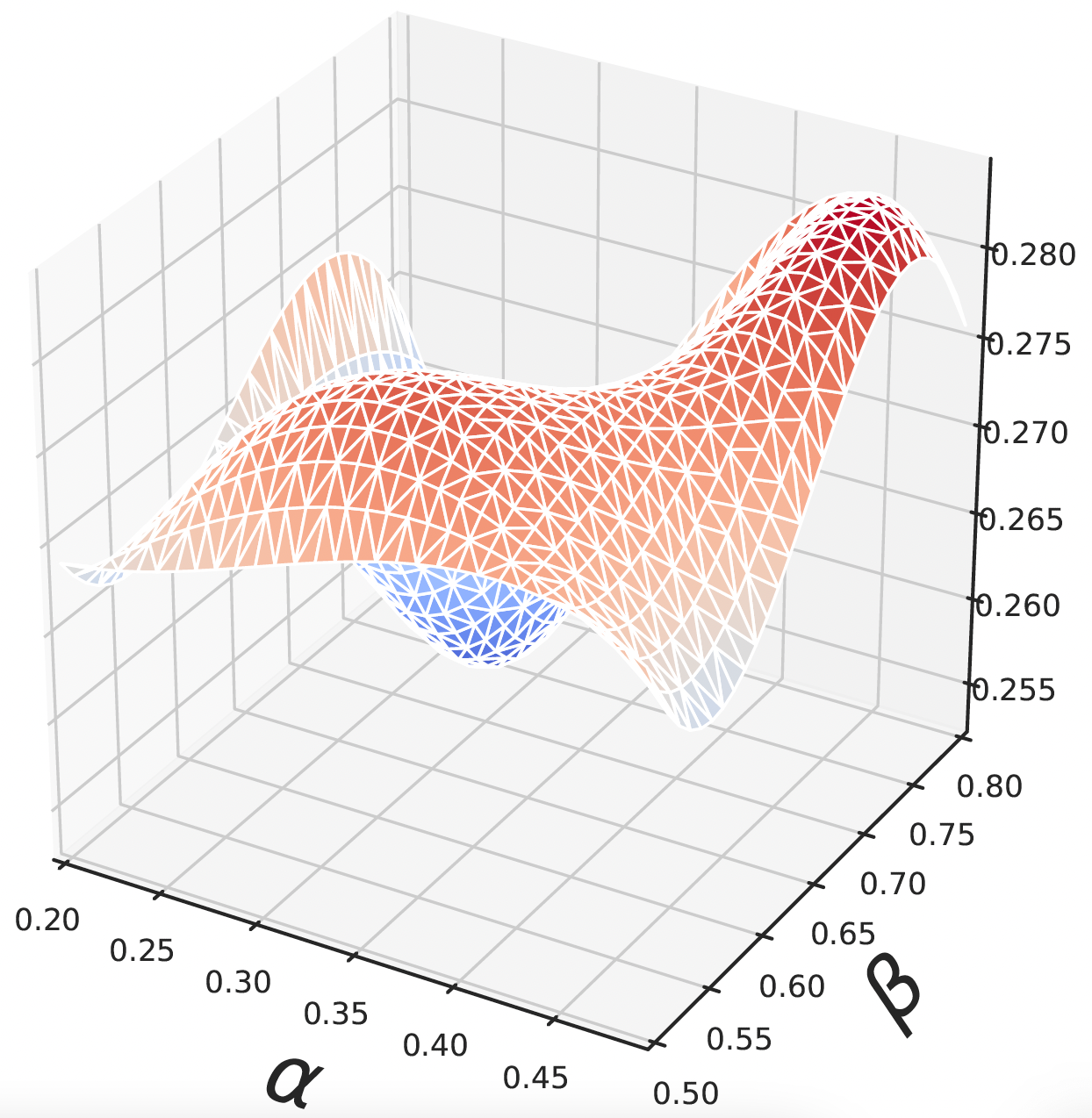}
}
\hspace{-1mm}

\subfigure[AUC of Data-A]{
\includegraphics[width=0.14\textwidth]{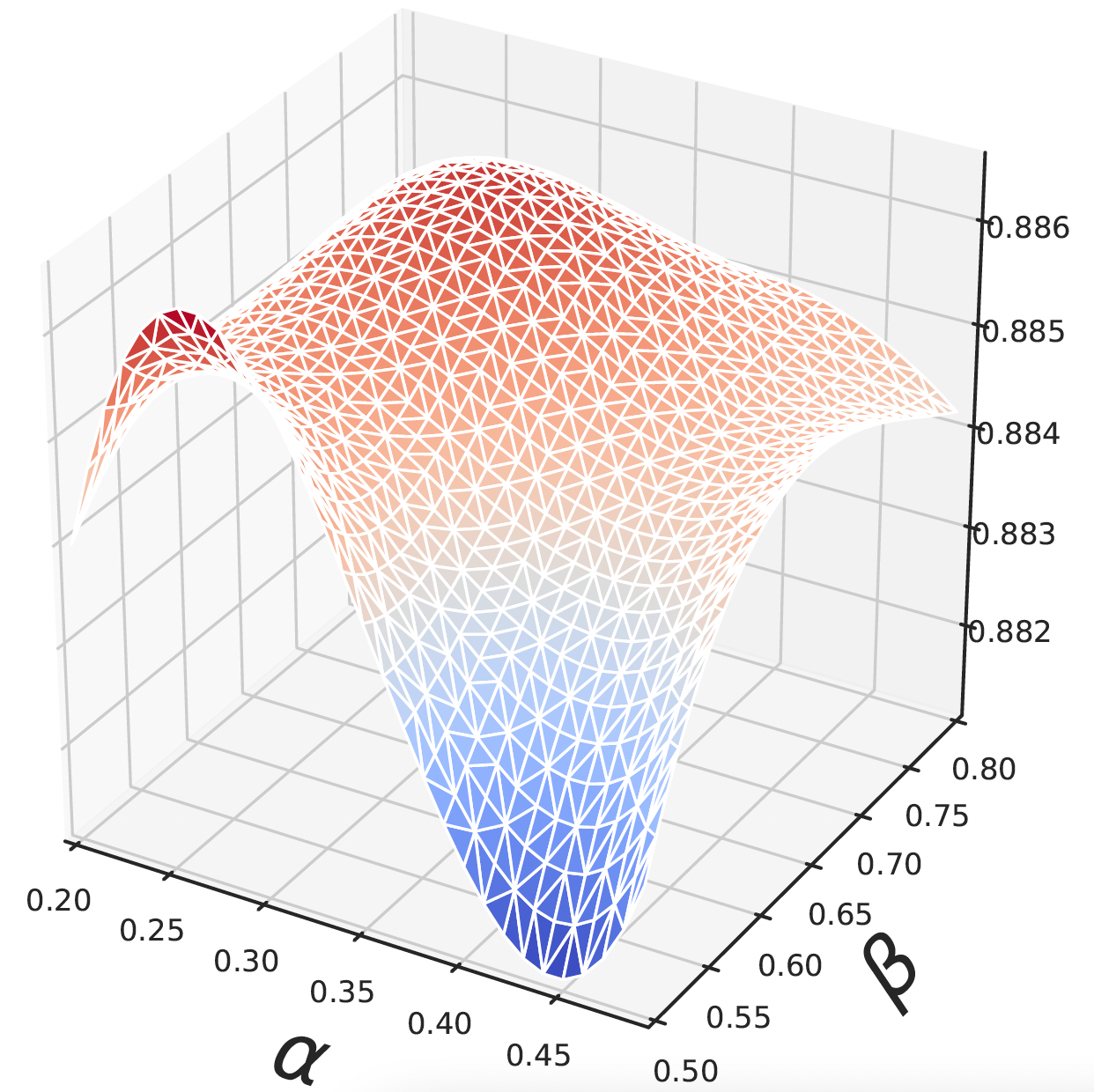}
}
\hspace{-1mm}
\subfigure[F1-score of Data-A]{
\includegraphics[width=0.14\textwidth]{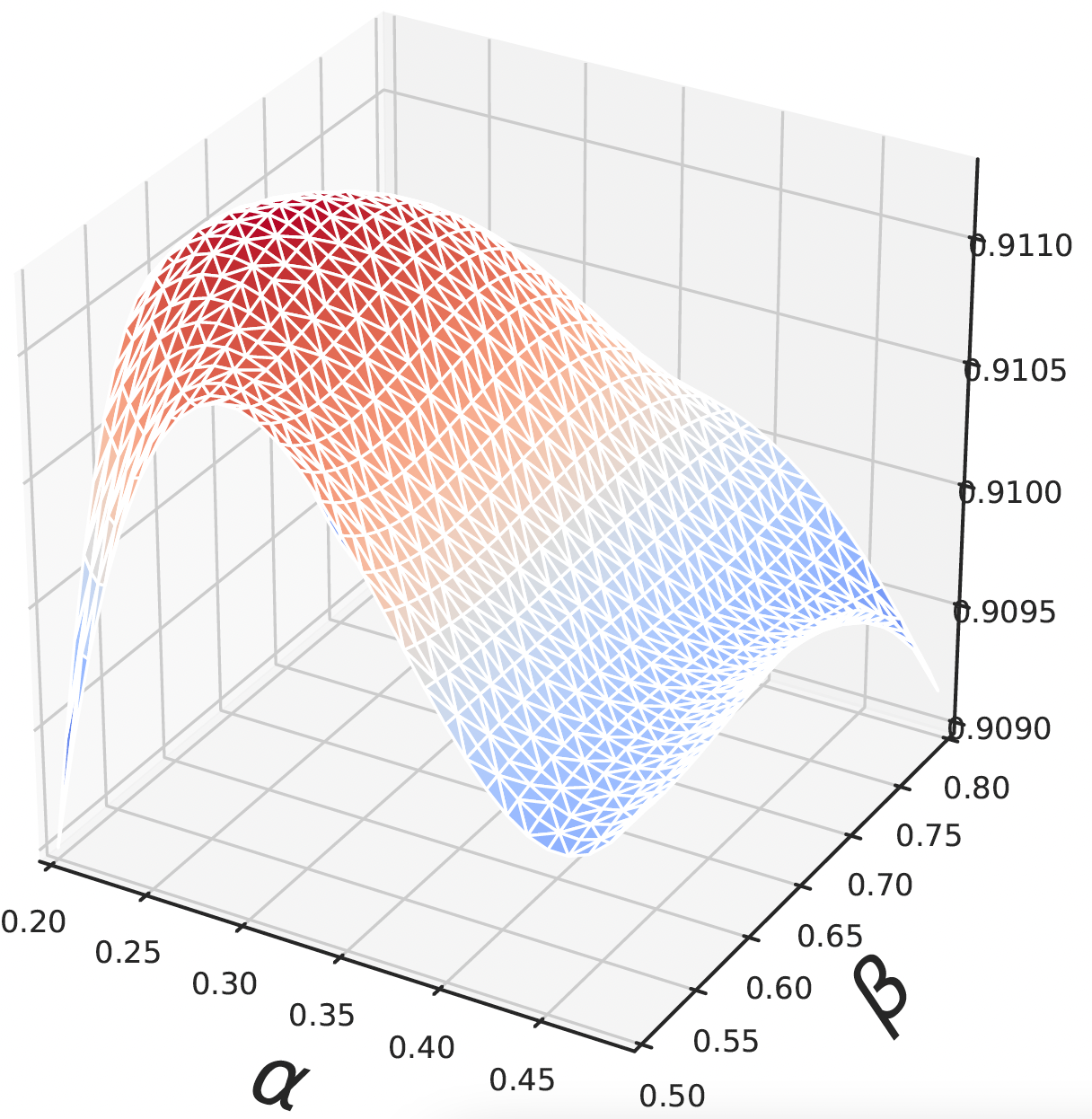}
}
\hspace{-1mm}
\subfigure[FNR of Data-A]{
\includegraphics[width=0.14\textwidth]{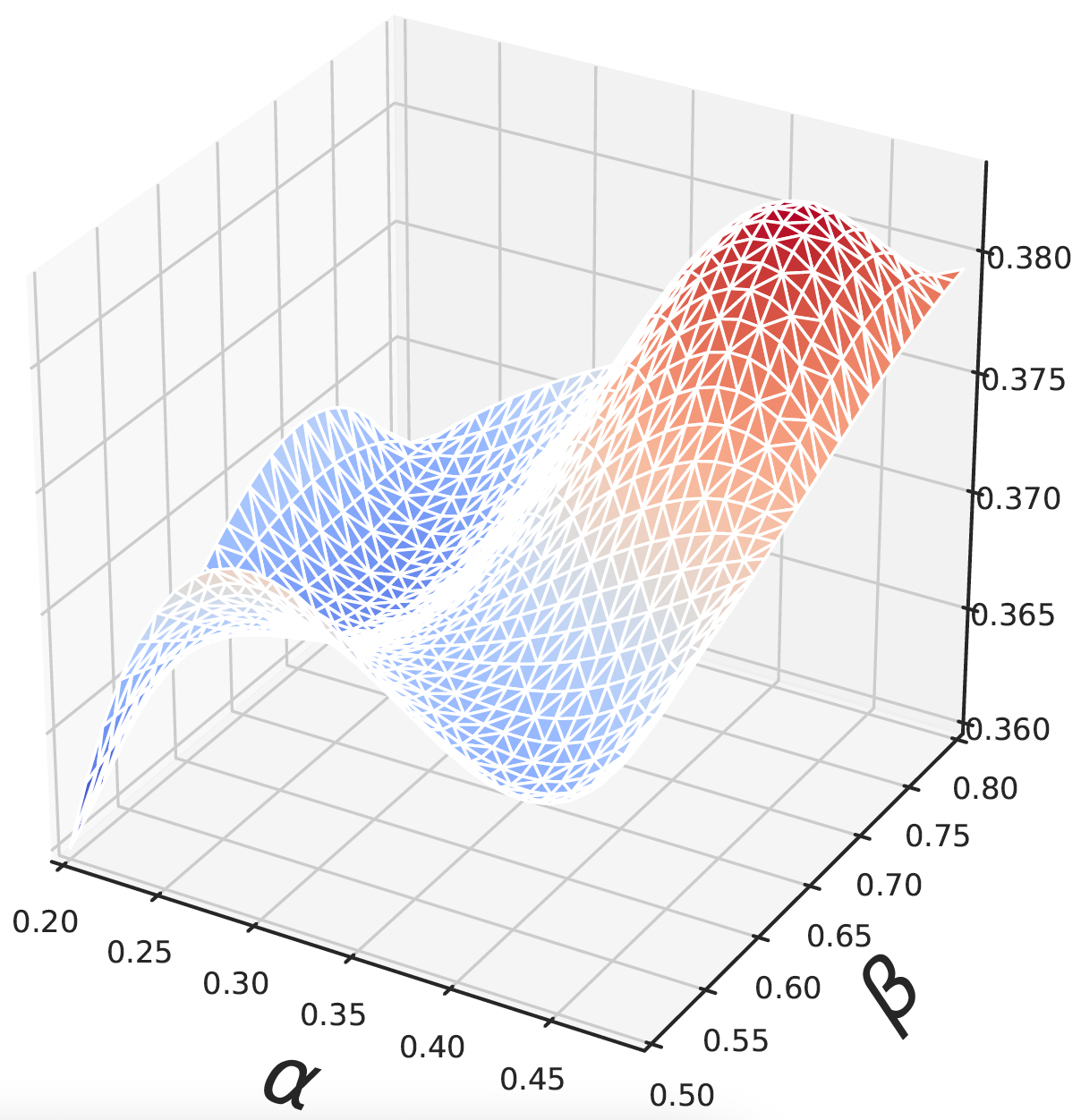}
}
\caption{Effect of different values of $\alpha$ and $\beta$. (a), (b), and (c) are the results of Data-E; (d), (e), and (f) are the results of Data-A. The red (blue) color corresponds to the high (low) value.}
\label{sensitivity-alpha}
\end{figure}

\begin{table}[ht]
\setlength\tabcolsep{3pt}
\scriptsize
\setlength{\abovecaptionskip}{-0cm} 
\setlength{\belowcaptionskip}{-0mm}
\caption{Effect of different neighbor selection strategies on Data-E.}
\begin{center}
\begin{tabular}{ccccccc}
\toprule[1.0pt]
\multirow{2}*{\small \bf Rate}   & \multicolumn{3}{c}{\small \bf Click+BERT's score} & \multicolumn{3}{c}{\small \bf Purchase+BERT's score}   \\
&\bf AUC &\bf F1-score &\bf FNR($\downarrow)$ &\bf AUC &\bf F1-score &\bf FNR($\downarrow$) \\ \midrule[0.5pt]
$\lambda=0.0$	&\textbf{0.8785} &\textbf{0.8256} &0.2966 &\textbf{0.8785}  &\textbf{0.8256} &0.2966  \\
$\lambda=0.2$	&0.8779 &0.8237 &\textbf{0.2834} &0.8777  &0.8236 &\textbf{0.2810}  \\
$\lambda=0.4$	&0.8770 &0.8200 &0.3198 &0.8756  &0.8214 &0.3068  \\
$\lambda=0.6$	&0.8670 &0.8085 &0.4000 &0.8696  &0.8045 &0.3694  \\
$\lambda=0.8$	&0.8679 &0.8101 &0.3901 &0.8660  &0.8105 &0.3262  \\
$\lambda=1.0$	&0.8656 &0.8091 &0.3850 &0.8671  &0.8109 &0.3727  \\
\bottomrule[1.0pt]
\end{tabular}
\end{center}
\label{neighbor1}
\end{table}

\begin{table}[ht]
\setlength\tabcolsep{3pt}
\scriptsize
\setlength{\abovecaptionskip}{-0cm} 
\setlength{\belowcaptionskip}{-0mm}
\caption{Effect of different neighbor selection strategies on Data-A.}
\begin{center}
\begin{tabular}{ccccccc}
\toprule[1.0pt]
\multirow{2}*{\small \bf Rate}   & \multicolumn{3}{c}{\small \bf Click+BERT's score} & \multicolumn{3}{c}{\small \bf Purchase+BERT's score}   \\
&\bf AUC &\bf F1-score &\bf FNR($\downarrow$) &\bf AUC &\bf F1-score &\bf FNR($\downarrow$) \\ \midrule[0.5pt]
$\lambda=0.0$  &\textbf{0.8862} &0.9107 &\textbf{0.3673} &\textbf{0.8862} &0.9107 &\textbf{0.3673} \\
$\lambda=0.2$	&0.8849 &\textbf{0.9113} &0.3794 &0.8830  &\textbf{0.9117} &0.3831  \\
$\lambda=0.4$	&0.8821 &0.9112 &0.4016 &0.8818  &0.9100 &0.4131  \\
$\lambda=0.6$	&0.8779 &0.9068 &0.4793 &0.8783  &0.9064 &0.4844  \\
$\lambda=0.8$	&0.8802 &0.9076 &0.4676 &0.8790  &0.9084 &0.4683  \\
$\lambda=1.0$	&0.8792 &0.9077 &0.4671 &0.8787  &0.9079 &0.4716  \\
\bottomrule[1.0pt]
\end{tabular}
\end{center}
\label{neighbor2}
\end{table}

\subsubsection{Thresholds $k$}
Here we evaluate the effects of $k$ on the performance of BERM by sampling neighboring nodes with different hops from the bipartite graph. The comparison results on Data-E and Data-S are shown in Fig.~\ref{different_k}. We can see that the BERM with $k$=2 achieves the best performance among them. When $k$ is too large such as $k=5$, many distant neighbors are aggregated into the anchor nodes, then it leads to the performance degradation of BERM due to lots of noise gathering in these distant neighbors. Therefore, we conclude that $2$-order relevance matching modeling is the optimal choice for our e-commerce scene.

\begin{figure}[t]
\centering
\includegraphics[width=0.35\textwidth]{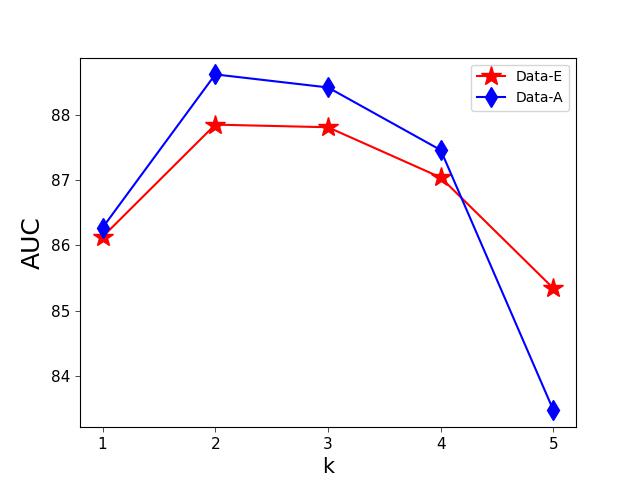}
\hspace{0in}
\caption{The effects of different $k$.}
\label{different_k}
\vspace*{-4mm}
\end{figure}

\subsubsection{Selection of Neighbor Structure}
In BERM, the selection of neighbor structure directly affects which context information is transmitted to the anchor node. A good selection strategy can aggregate valuable neighboring node information to enrich the anchor node's representation. To investigate the effect of different neighbor structure selection strategies on BERM and seek a relatively optimal solution, we use different values of hyper-parameter $\lambda$ to control the ratio between user behavior and BERT's score.
Specifically, we calculate a new score $Score_{new}(Q, I) = \lambda * User(Q, I) + (1-\lambda) * Score(Q, I) $ where $User(Q, I)$ is the user behavior feature (e.g., for click behavior, $User(Q, I) = 1$ if click behavior happens between query $Q$ and item $I$). The addition and deletion of edges refer to $Score_{new}(Q, I)$, rather than $Score_(Q, I)$.
We report the results with different $\lambda$ in Tab.~\ref{neighbor1} and~\ref{neighbor2}. From them, we can conclude that: 
\begin{itemize}[leftmargin=*]
\item Using BERT's score is better than using user behaviors for the selection of neighbors. 
Therefore, the value of AUC or F1-score gradually decreases with the increase of $\lambda$; the value of FNR increases with the increase of $\lambda$. The optimal $\lambda\in \left [0.0, 0.2\right ]$.
\item According to the metric of FNR, the purchase behavior is better than the click behavior on Data-E. The reason for it is that purchase behaviors reveal more accurate semantic relevance information than click behaviors. However, the purchase behavior is worse than the click behavior on Data-A. We think that it is caused by the sparsity of purchase behaviors in the dataset of all categories.
\end{itemize}

\end{document}